# Analysis of the hierarchical structure of the *B. subtilis* transcriptional regulatory network

Santhust Kumar,[a] Michele Vendruscolo,[b] Amit Singh,[c] Dhiraj Kumar,[d] and Areejit Samal[*ef]

[a] Department of Physics and Astrophysics, University of Delhi, Delhi India
[b] Department of Chemistry, University of Cambridge, Cambridge, UK
[c] Department of Microbiology and Cell Biology, Indian Institute of Science, Bangalore, India
[d] International Centre for Genetic Engineering and Biotechnology, New Delhi, India
[e] International Centre for Theoretical Physics, Trieste, Italy
[f] The Institute of Mathematical Sciences, Chennai, India
*asamal@ictp.it

The transcriptional regulation of gene expression is orchestrated by complex networks of interacting genes. Increasing evidence indicates that these 'transcriptional regulatory networks' (TRNs) in bacteria have an inherently hierarchical architecture, although the design principles and the specific advantages offered by this type of organization have not yet been fully elucidated. In this study, we focussed on the hierarchical structure of the TRN of the gram-positive bacterium *Bacillus subtilis* and performed a comparative analysis with the TRN of the gram-negative bacterium *Escherichia coli*. Using a graph-theoretic approach, we organized the transcription factors (TFs) and σ-factors in the TRNs of *B. subtilis* and *E. coli* into three hierarchical levels (Top, Middle and Bottom) and studied several structural and functional properties across them. In addition to many similarities, we found also specific differences, explaining the majority of them with variations in the distribution of σ-factors across the hierarchical levels in the two organisms. We then investigated the control of target metabolic genes by transcriptional regulators to characterize the differential regulation of three distinct metabolic subsystems (catabolism, anabolism and central energy metabolism). These results suggest that the hierarchical architecture that we observed in *B. subtilis* represents an effective organization of its TRN to achieve flexibility in the response to diverse stimuli.

## Introduction

Bacteria adapt to environmental changes by tuning their gene expression in response to external and internal stimuli. This process occurs primarily through transcriptional regulation, which involves context-specific binding of transcriptional regulators upstream of target gene sequence. Transcriptional control is exercised through regulators, including transcription factors (TFs) and σ-factors, which are themselves subject to transcriptional regulation. Thus, transcriptional regulation is achieved through a directed network of interacting genes – the transcriptional regulatory network (TRN)[1-9] – where nodes represent genes (regulators or targets) and directed edges represent regulatory interactions signifying transcriptional control of target gene expression by regulators. A major goal of systems biology is to elucidate the design principles[2, 3, 5, 10-13] governing the global organization of TRNs.

The description of transcriptional regulatory interactions in the language of directed networks has provided novel insights on the structural organization of TRNs using methods developed to analyze complex networks[3, 5, 11, 13, 14]. It has thus been realised that there is a broad distribution[5, 11, 15] in the number of target genes directly regulated by a TF, and there are repeated occurrences of certain subgraphs, known as 'network motifs'[3, 16] in TRNs. Several studies on the large-scale structure of TRNs, including in particular *Escherichia coli* and *Saccharomyces cerevisiae*, have established the existence of an inherent hierarchical architecture with limited feedback loops[8, 9, 14, 17-20]. The hierarchical structure of the TRN of *E. coli* has also been shown to enable cellular homeostasis and flexibility of responses to environmental changes[18]. This architecture of TRNs allows the organization of transcriptional regulators and target genes into different levels[8, 9, 14, 17-20]. Investigations mainly in *E. coli*[8, 9, 17, 18, 20] and *S. cerevisiae*[9, 19, 20] have shown that genes in different hierarchical levels of TRNs have distinct structural, dynamical and evolutionary properties.

In this work, we studied the hierarchical structure of TRN in the gram-positive bacterium *Bacillus subtilis* and investigated which aspects in the hierarchical structure of its TRN are more important to determine the responses to environmental stimuli. To this end, we compared the TRN of *B. subtilis* with that of the gram-negative bacterium *E. coli*, which has the best characterized TRN to date. *B. subtilis* and *E. coli* are bacteria with similar genome sizes that have diverged more than one billion years ago. *B. subtilis* is a free living bacterium commonly found in soil but that has the ability to grow in diverse environments, from the gastrointestinal tract to the root surface of plants, while *E. coli* is commonly found in the gut of warm-blooded higher organisms. Thus, *B. subtilis,* in contrast to *E. coli,* has a lifestyle that exposes it to many more uncertainties in the form of diverse, and sometimes extreme, environmental conditions. *B. subtilis* can adapt to such conditions, which include stress and nutrient limitation, through sporulation which is associated with distinct regulatory programs[21], while *E. coli* is not known to sporulate.

Despite having similar genome sizes, one feature in which the TRNs of *B. subtilis* and *E. coli* differ significantly is the number of σ-factors, which are proteins that help regulate transcription initiation of specific genes by enabling the recruitment of the transcriptional machinery. Thus, σ-factors impose an additional layer of regulation in gene expression because of their selectivity in binding to different gene promoters[22]. *B. subtilis* has twice as many σ-factors as *E. coli*, which may reflect the necessity of *B. subtilis* to have a broad range of regulatory mechanisms to cope with greater uncertainties in its environment. To understand the significance of σ-factors in shaping the organization of TRNs, we thus compared the structural and functional properties of *B. subtilis* and *E. coli* TRNs with and without the inclusion of σ-factors.

We considered the most recent reconstructions of the TRNs of *B. subtilis*[23] and *E. coli*[24]. By analysing a series of recently proposed graph-theoretic measures[25] we quantified the extent of hierarchical organization in the TRNs of the two organisms studied here. Using well-established graph-theoretic algorithms[9, 19, 20], we next classified transcriptional regulators into different hierarchical levels and studied the enrichment of various structural and functional properties in different levels of hierarchy in the two organisms. Our study reveals many unifying features, as well as some distinct ones, in the enrichment of structural and functional properties in different hierarchical levels of the TRNs of *B. subtilis* and *E. coli*. Our results thus complement those of a recent study[26] in which the role of gene duplication and divergence in shaping the hierarchical structure of TRNs in *B. subtilis, E. coli* and yeast was investigated.

## Results and Discussion

### *B. subtilis* and *E. coli* transcriptional regulatory networks with and without σ-factors

We compared the TRN of *B. subtilis,* which comprises 1594 (protein coding) genes and 2976 interactions obtained from reconstruction by Freyre-Gonzalez *et al*[23] with the TRN of *E. coli,* which contains 3073 (protein coding) genes and 7977 interactions extracted from RegulonDB[24] database (see Methods and Table S1). Since the TRN of *E. coli* is very well characterized, it is not surprising that the number of known interactions and target genes in the TRN of *B. subtilis* is approximately half of that in the TRN of *E. coli* (Table S1). Although the level of characterization of the TRN of *B. subtilis* is lower than that of the TRN of *E. coli*, the density of edges in the TRNs of the two organisms is similar (Table S1). We can thus expect that the statistics of density of edges may not change very much even as number of known interactions and target genes in the TRN of *B. subtilis* will increase through future studies.

One important aspect of transcriptional regulation in which *B. subtilis* and *E. coli* differ significantly is in the number of σ-factors. *B. subtilis* has twice the σ-factors compared to *E. coli* (14 in *B. subtilis* to 7 in *E. coli*). This difference is consistent with the idea that *B. subtilis* needs a broad range of regulatory mechanisms to cope with uncertainties in its environment.

We investigated the role played by σ-factors in organization of TRNs by comparing the structural and functional properties of the TRNs of *B. subtilis* and *E. coli* with and without σ-factors (see Methods). We found that the exclusion of σ-factors from the TRNs of *B. subtilis* and *E. coli* results in a significant decrease in the number of regulatory interactions and in the clustering coefficient of the TRNs (Table S1).

**Feedback processes in transcriptional regulatory networks**

As feedback processes in TRNs indicate departure from a strict hierarchical structure, we quantified the amount of feedback in the TRNs of *B. subtilis* and *E. coli* (with and without σ-factors) by measuring the size of the largest strongly connected component (LSCC). A strongly connected component (SCC) within a directed graph is a maximal set of nodes such that for any pair of nodes *i* and *j* in the set there is a directed path from *i* to *j* and from *j* to *i*. Thus, any SCC is a cycle in the directed graph.

The size of LSCC in the TRN of *B. subtilis* is smaller than that in the TRN of *E. coli* (Table S1). Crucially, the size of the LSCC in the TRNs of each organism increases by more than three times when σ-factors are included in the TRNs (Table S1). Thus, the inclusion of σ-factors increases not just the connectivity but also the amount of feedback in TRNs of both organisms (Table S1). However, by comparing the size of the LSCCs in *B. subtilis* and *E. coli* against networks that were randomized in a manner that preserved the in-degree and out-degree at each gene, we found that the size of the LSCC in each organism is much smaller than expected by chance (Table S1). These results indicate that the TRNs of *B. subtilis* and *E. coli* exhibit limited feedback compared to the corresponding randomized networks.

We also studied the Perron-Frobenius eigenvalue associated with the LSCC, which provides a measure of the multiplicity of pathways within the cycle. We found that the Perron-Frobenius eigenvalue of the TRN of *B. subtilis* is smaller than that of the TRN of *E. coli,* and that its value increases with the inclusion of σ-factors (Table S1).

In the case of *E. coli*, the number of known regulatory interactions in RegulonDB[24, 27] has grown by more than tenfold in the last 15 years, leading to an increase in the density of edges and in the size of the LSCC. However, the size of the LSCC has consistently remained smaller than expected for a randomized network. Based on these trends in *E. coli* we may expect that, although future expansion in the TRN of *B. subtilis* could lead to an increase in the size of its LSCC, the amount of feedback should remain smaller than expected in the corresponding randomized networks.

**Hierarchical organization of transcriptional regulatory networks**

The results discussed above are consistent with those of recent studies, which established that the global structure of TRNs in microorganisms can be characterized by a largely hierarchical structure[8, 9, 14, 17-20] with limited feedback in transcriptional regulation. We next quantified the extent of hierarchical organization in the TRNs of *B. subtilis* and *E. coli* and classified their genes into different levels of hierarchy.

Recently Corominas-Murtra *et al*[25] proposed three measures, Treeness (T), Feedforwardness (F) and Orderability (O), to quantify the extent of hierarchical organization in complex directed networks. In a given network, the treeness quantifies the extent of pyramidal structure and unambiguity in the chain of command, the feedforwardness measures the impact of feedback processes in the casual flow of information, and the orderability gives the fraction of nodes that does not belong to any cycle. We computed these three measures for the TRNs of *B. subtilis* and *E. coli.* Based on the T, F and O values that we obtained, we concluded that the TRNs of two organisms have a largely hierarchical structure (Table S1). The values for the TRN of *B. subtilis* were similar to those obtained for the TRNs of other organisms by Corominas-Murtra *et al*[25].

An important factor governing the timely response of TRNs to environmental changes is represented by the number of levels in their hierarchical organization. Using a vertex-sort algorithm[19] we determined the number of levels in the Top-down and Bottom-up hierarchical decomposition of the TRNs of *B. subtilis* and *E. coli* (see Methods). The number of levels was found to be smaller than that observed in randomized networks (Table S1). Hence, the TRNs of *B. subtilis* and *E. coli* display limited depth in their hierarchical structure suggesting a possible dynamical optimization in the regulation of targets[17, 18]. These results are consistent with those by Sellerio *et al*[26] who used a different hierarchical decomposition method and earlier versions of the TRNs of *B. subtilis* and *E. coli*.

After establishing that the TRNs of *B. subtilis* and *E. coli*, have a largely hierarchical organization, we classified the transcriptional regulators in the two organisms into a three-level hierarchy: Top, Middle and Bottom (see Methods and Table S2). Based on this classification, we found that the TRNs of *B. subtilis* and *E. coli* with σ-factors have a pyramidal structure (Table S2 and Figure 1). This organization may reflect an optimization for effecting large downstream changes by controlling few regulators upstream in the hierarchical structure of TRNs (Figure 1).

**Enrichment of structural and functional properties in different levels of hierarchy in transcriptional regulatory networks**

*Hubs*

The out-degree of a transcriptional regulator in a given TRN gives the number of genes directly regulated by it. Earlier studies have established that the out-degree distribution for transcriptional regulators in the TRNs of *B. subtilis*[23] and *E. coli*[3, 16] follows a power law[28] where most regulators have low out-degree while few regulators (referred to as 'hubs') have very high out-degree. Hubs have been shown to be critical for the maintenance of the large-scale structure of complex networks[11, 29]. We studied the average out-degree and distribution of hubs in different levels of hierarchy in the TRNs of *B. subtilis* and *E. coli* (with and without σ-factors), finding that the Top and Middle levels have higher average out-degree and are enriched in hubs (Figure 2A,B). The average out-degree is highest for Middle level transcriptional regulators in all cases except for the TRN of *B. subtilis* with σ-factors. (Figure 2A). Hence, Middle level regulators control many downstream target genes in the TRN and are highly influential. Our results are consistent with those obtained for *E. coli* and yeast by Yu *et al*[9] and Jothi *et al*[19].

The exception in the case of the TRN of *B. subtilis* with σ-factors can be explained via comparison with the TRN of *E. coli* with σ-factors. In *B. subtilis* σ-factors are scattered across all three levels of hierarchy whereas in *E. coli* almost all σ-factors are in the Middle level (Figure 3). Of the σ-factors in *B. subtilis* and *E. coli*, RpoD has maximum number of targets in both organisms. In *B. subtilis*, RpoD accounts for almost half of the edges and in *E. coli* almost a third of the edges in the network. However, RpoD is located in the Top level in *B. subtilis* while being in Middle level in *E. coli* (Figure 3). A future growth in the number of known interactions in the TRN of *B. subtilis* may result in the possible addition of edges associated with RpoD and other σ-factors, which in turn may lead to a universal conclusion at that juncture.

*Bottlenecks*

An efficient transmission of information in the TRNs is critical for achieving timely and appropriate responses to external and internal stimuli. Bottlenecks in TRNs correspond to genes through which many shortest paths pass and are important for efficient flow of information. The betweenness centrality[30, 31] is a graph-theoretic measure that quantifies the number of shortest paths passing through a node in the network. Thus, bottlenecks are defined as nodes with high betweenness centrality. We studied the average betweenness centrality and distribution of bottlenecks in different levels of hierarchy in the TRNs of *B. subtilis* and *E. coli* (with and without σ-factors), and found that the Middle level has the highest betweenness centrality and are enriched in bottleneck regulators in both organisms (Figure 2C,D). These results are consistent with that obtained for *E. coli* and yeast by Yu *et al*[9]. Hence, the information flow from Top level regulators to target genes predominantly passes through Middle level regulators in the TRNs of *B. subtilis* and *E. coli*.

*Coregulation of genes by transcriptional regulators*

Bhardwaj *et al*[20] proposed two measures to quantify coregulatory partnerships between transcription regulators, the degree of collaboration and the degree of pair collaboration. The degree of collaboration of a transcriptional regulator measures the faction of target genes that are coregulated by at least one other regulator. We studied the average degree of collaboration for regulators in different levels of hierarchy in the TRNs of *B. subtilis* and *E. coli* (with and without σ-factors), and found that the Middle level regulators are more collaborative than regulators in other levels (Figure 2E). The degree of pair collaboration for a pair of transcriptional regulators measures the number of genes coregulated by the pair divided by the number of genes regulated by at least one of the regulators in the pair. We used this measure to quantify the extent of intra- and inter-level pair coregulatory partnerships of and between different levels of hierarchy in the TRNs of *B. subtilis* and *E. coli* (with and without σ-factors). We found that the average degree of pair collaboration is highest for pair of regulators from the Middle level (that is, the Middle-

Middle) followed by pair of regulators where one regulator belongs to the Top level and other belongs to the Middle level (that is, the Top-Middle) in all cases except for the TRN of *E. coli* with σ-factors (Figure 4). Our results, especially for the TRN of *B. subtilis*, match those obtained by Bhardwaj et al[20] and Jothi et al[19] for other organisms where the Middle-Middle had the highest propensity for pair collaboration (Figure 4).

*Evolutionary conservation of transcriptional regulators*

The evolutionary conservation of transcriptional regulators in distant organisms can be studied through orthologous genes. We thus extracted the list of orthologous genes in *B. subtilis* and *E. coli* from the KEGG[32, 33] database (see Methods), and then studied the evolutionary conservation of regulators in different levels of hierarchy in their TRNs. We found that the Top level transcriptional regulators are more conserved between the two bacteria compared to the Middle and Bottom level regulators (Figure 2F). These results may indicate that general transcription factors are more conserved between these two distant bacteria, consistent with what was reported by Jothi et al[19] in the case of yeast.

*Feed Forward Loops*

Feed forward loops (FFLs) are network motifs that commonly occur in TRNs[3]. FFL is a 3-node subgraph (circuit) composed of regulator X, regulator Y and target gene Z. In FFL, X regulates Y and Z, while Y regulates Z. FFL motifs have been shown to perform important dynamical functions[34] in TRN. We studied the composition of FFLs based on genes from different levels of hierarchy in the TRNs of *B. subtilis* and *E. coli* (with and without σ-factors). Top level regulators along with Bottom level regulators appear more often in FFLs in the TRN of *B. subtilis* while Middle level regulators appear more often in FFLs in the TRN of *E. coli* (Figure 5 and Table S3). Dissimilarity in FFL composition in the TRNs of *B. subtilis* and *E. coli* (Figure 5) can be explained by differences in number of inter- and intra-level edges between levels of the TRNs in the two organisms (Table S2). In the TRN of *B. subtilis* most edges are between Top level regulators and Target genes while in the TRN of *E. coli* most edges are between Middle level regulators and Target genes.

*Two-component Regulatory Systems*

Two-component regulatory systems are basic stimulus-response systems in prokaryotes for sensing environmental changes, which are typically composed of a sensory kinase and a response regulator[35]. We studied the distribution of two-component system genes in the different levels of the hierarchy in the TRNs of *B. subtilis* and *E. coli* with and without σ-factors, and found that the Top and Middle levels of hierarchy are enriched in two-component system regulators (Figure 6). Preponderance of two-component system regulators in the Top and Middle levels indicate that regulators responding to environmental changes lie upstream in the hierarchy of the TRNs of *B. subtilis* and *E. coli*.

**Regulation of distinct metabolic subsystems by *B. subtilis* and *E. coli* transcriptional regulatory networks**

Up to this point we mainly focussed on structural and functional properties of transcriptional regulators in different levels of hierarchy in the TRNs of *B. subtilis* and *E. coli*. We next investigated the regulation of target genes coding for enzymes in distinct metabolic subsystems by the TRNs of *B. subtilis* and *E. coli*. For this analysis, we used pathway information in Metacyc[36] database to classify target genes coding for enzymes in the TRNs of *B. subtilis* and *E. coli* into three broad biochemical categories: Catabolism, Anabolism and Central Energy Metabolism (see Methods and Table S4). Catabolic enzymes are responsible for the uptake of nutrient molecules from the environment and their breakdown into simpler metabolites that feed into central metabolism. Anabolic enzymes are responsible for synthesis of biomass components from precursor metabolites required for growth. Central energy metabolism enzymes are situated between catabolism and anabolism, and are responsible for generating energy and precursor metabolites.

We determined the number of transcriptional regulators (TFs and σ-factors, separately) controlling target genes coding for enzymes in the three distinct metabolic subsystems in *B. subtilis* and *E. coli* (Figure 7 and Table S5). We did not find differences in the average number of σ-factors controlling target genes coding for enzymes in the three distinct metabolic subsystems in the two organisms (Figure 7 and Table S5). Hence, the three distinct metabolic subsystems (catabolism, anabolism and central energy metabolism) do not appear to be differentially regulated by σ-factors in the two organisms. However, we did find difference in the average number of TFs controlling target genes coding for enzymes in the three distinct metabolic subsystems in both organisms (Figure 7 and Table S5). The average number of TFs controlling target genes coding for anabolic enzymes is very low in both *B. subtilis* and *E. coli* (Figure 7 and Table S5). Thus, anabolism is least tightly regulated in both organisms. In *B. subtilis*, the average number of TFs controlling catabolic enzymes is higher than that for central energy metabolism enzymes, while in *E. coli*, the average number of TFs controlling catabolic enzymes is lower than that for central energy metabolism enzymes (Figure 7 and Table S5). Thus, in both organisms, catabolic and central energy metabolism enzymes are more tightly regulated than anabolic enzymes.

Our analysis of regulation of distinct metabolic subsystems in *B. subtilis* and *E. coli* was inspired by similar investigation by Seshasayee et al[37] in *E. coli*. Seshasayee et al[37] use the TRN of *E. coli* from an earlier version of RegulonDB[38] for their study while we used the latest version of RegulonDB[24]. However, consistently with Seshasayee et al[37] we found that in *E. coli*, anabolic enzymes are least regulated by TFs, followed by catabolic enzymes and then by central energy metabolism enzymes (Figure 7 and Table S5).

We found that the regulation of three distinct metabolic subsystems in *B. subtilis* and *E. coli* do not match in the order for catabolic and central energy metabolism enzymes (Figure 7 and Table S5). In *B. subtilis*, the average number of TFs controlling catabolic enzymes is slightly higher than that for central energy metabolism enzymes. However, in *E. coli*, the average number of TFs controlling catabolic enzymes is much less than that for central energy metabolism enzymes. Since the TRN of *B. subtilis* and its metabolism are much less characterized than those of *E. coli*, it is possible that future expansion in the TRN of *B. subtilis* may lead to a different conclusion. Based on this analysis, we can also advise future curators of the TRN of *B. subtilis* to strategically focus on filling knowledge gaps in regulation of central energy metabolism genes.

## Conclusions

We have compared the hierarchical structure of the transcriptional regulatory networks (TRNs) of two evolutionarily distant bacteria, *B. subtilis* and *E. coli*, which have similar genome sizes but different life styles. We have first determined the extent of the hierarchical organization of the TRNs using a range of recently proposed measures, including Treeness, Feedforwardness and Orderability[25]. We have then combined decomposition approaches[19, 20] to classify the transcriptional regulators in the TRNs of *B. subtilis* and *E. coli* into three distinct hierarchical levels (Top, Middle and Bottom), and studied in detail the enrichment of several structural and functional properties across them.

A novel aspect of this study is represented by the use of σ-factors to dissect their role in determining the architecture of the TRNs of *B. subtilis* and *E. coli*. One could expect that a network without σ-factors would be mostly context-independent with loss of selectivity and specificity brought via σ-factors in the complete network. Yet, we have found that even without σ-factors the TRNs of the two organisms that we considered largely retain several of the structural and functional features studied here. We have also found, however, that the dissimilarities in the enrichment of specific properties can be explained by differences in the distributions of σ-factors across the hierarchical levels in the two organisms.

Our study of two evolutionary distant bacteria therefore underscores the universality in the design principles of bacterial regulatory networks by identifying some aspects of the large-scale organization of TRNs into inherent hierarchical structures where transcriptional regulators across different hierarchical levels have distinct structural and functional properties. Taken together these results suggest that the observed hierarchical architecture of TRN may represent a very effective organization for transcription regulation even when bacteria need to respond to only limited stimuli.

## Methods

**Datasets**

*Transcription Regulatory Network*
The TRN of *B. subtilis* was obtained from the recent reconstruction by Freyre-Gonzalez *et al*[23] which is a curated database of regulatory interactions with strong evidence from DBTBS version 2010[39]. In this work, we excluded the ncRNA (e.g., sRNA, tRNA, rRNA, misc_RNA) and their regulatory interactions from TRNs. After excluding ncRNA and their interactions from the Freyre-Gonzalez *et al*[23] reconstruction, we obtained a TRN of *B. subtilis* with 140 transcriptional regulators (126 TFs, 14 σ-factors), 1594 (protein coding) genes and 2976 interactions (Tables S1). The TRN of *E. coli* was extracted from RegulonDB[24] database. After excluding ncRNA and their interactions in RegulonDB[24], we obtained a TRN of *E. coli* with 202 transcriptional regulators (195 TFs, 7 σ-factors), 3073 (protein coding) genes and 7977 interactions (Table S1). We converted the common names of protein coding genes in the TRNs of *B. subtilis* and *E. coli* to their unique numeric identifiers, BSU- and b- numbers, respectively.

An important aspect of this study is to investigate the role played by σ-factors in organization of TRNs in *B. subtilis* and *E. coli*. Hence, we studied the TRNs of *B. subtilis* and *E. coli* with and without σ-factors. The TRN of *B. subtilis* without σ-factors contains 126 TFs, 1054 genes and 1478 interactions and the TRN of *E. coli* without σ-factors has 195 TFs, 1643 genes and 4155 interactions (Table S1). Regulatory interactions in the TRNs of *B. subtilis* and *E. coli* with and without σ-factors are available in Tables S6-S9.

*Orthologous genes*
Orthologous genes in different species are genes that have descended from a common ancestral sequence and are a signature of evolutionary conservation. We extracted the list of orthologous genes in *B. subtilis* and *E. coli* genome from KEGG[32, 33] database.

*Two-component regulatory systems*
Two-component regulatory systems are mostly composed of a sensory kinase and a response regulator[35]. We compiled the set of known two-component regulatory systems in *B. subtilis* and *E. coli* from primary literature and several publicly accessible databases including P2CS[40], KEGG[32, 33], and Subtiwiki[41]. Our list of known two-component regulatory systems accounted for 75 and 63 genes in *B. subtilis* and *E. coli*, respectively.

*Classification of target genes into different metabolic subsystems*
Metacyc[36] database has classified genes in different organisms including those in *B. subtilis* and *E. coli* into different pathways. Metacyc[36] has grouped different metabolic pathways into three broad categories, namely, "Degradation/Utilization/Assimilation", "Biosynthesis" and "Generation of Precursor Metabolites and Energy". We used metabolic pathways in Bsubcyc[36] and Ecocyc[36] within Metacyc to classify enzyme coding target genes in the TRNs of *B. subtilis* and *E. coli* into the three broad categories that correspond to Catabolism, Anabolism, and Central Energy Metabolism. We excluded enzyme coding genes that appear in multiple categories (Table S4).

**Hierarchical decomposition of transcriptional regulatory networks**
We obtained the hierarchical decomposition of the TRNs of *B. subtilis* and *E. coli* into different levels as follows. At first, we determined genes with no outgoing edges in the directed graph associated with the TRN and assign them as target (TG) genes. Target genes predominantly code for metabolic enzymes. We then excluded target genes along with their edges from the TRN to obtain the key smaller network containing only interactions among transcriptional regulators[9, 19, 20]. We then identified strongly connected components (SCCs) in the directed graph associated with the smaller network containing only interactions among transcriptional regulators and collapse each SCC into a super node. The edges to (from) the genes in each SCC in the network are replaced by edges to (from) the corresponding super node to obtain a directed acyclic graph (DAG). Following Bhardwaj et al[20], we then classified the transcriptional regulators in the DAG into three levels based on connectivity: Nodes with no incoming edges (except self-regulation) in the DAG were assigned to the Top (T) level, nodes with no outgoing edges (except self-regulation) in the DAG were assigned to the Bottom (B) level, and the remaining nodes with both incoming and outgoing edges in the DAG were assigned to the Middle (M) level. Hence, the hierarchical decomposition of TRN classifies genes into four different levels: Top (T), Middle (M), Bottom (B) and Targets (TG) with first three levels corresponding to transcriptional regulators (Table S2). Note that our method of hierarchical decomposition of TRN into the four different levels differs from that followed by Bhardwaj et al[20] in following respect. Bhardwaj et al[20] do not construct DAG before assigning nodes to the Top, Middle and Bottom levels. However, we followed Jothi et al[19] to construct DAG before assigning nodes to the Top, Middle and Bottom levels. Hence, we allowed the possibility of genes in SCC to be assigned to the Top, Middle and Bottom levels in contrast to Bhardwaj et al[20].

We also applied the vertex-sort algorithm[19,42] to determine the number of actual levels in the TRNs of *B. subtilis* and *E. coli*. Leaf-removal procedure within the vertex-sort algorithm[19,42] can be used to decompose nodes into different levels in two different ways: Top-down and Bottom-up hierarchy. We determined the number of actual levels in both the Top-down and Bottom-up hierarchal decompositions of the TRNs of *B. subtilis* and *E. coli* (Table S1).

**Statistical significance**

To reveal the enrichment of specific properties (e.g. hubs, bottlenecks, degree of collaboration) of transcriptional regulators at different levels of hierarchy in *B. subtilis* and *E. coli*, we compared the value for the TRNs that we studied against randomized counterparts which preserve in- and out-degree at each gene in the network. The expected value of given properties of transcriptional regulators at different levels of hierarchy for randomized networks is shown as a dashed black line in our figures (Figure 2). For some properties (e.g. composition of FFLs), we reported also the Z-score to quantify the level of significance based on the comparison between values in the TRNs against the mean values and standard deviations in their randomized counterparts.

**Acknowledgements**

We thank A. Celani, S. Jain and M. Marsili for discussions. SK acknowledges The Institute of Mathematical Sciences for hospitality, University Grants Commission (UGC) India for Senior Research Fellowship, and University of Delhi grant DRCH/R&D/2013-14/4155 for infrastructural support.

**Supplementary Information**

Supplementary Tables S1-S9 are available upon request from the authors.

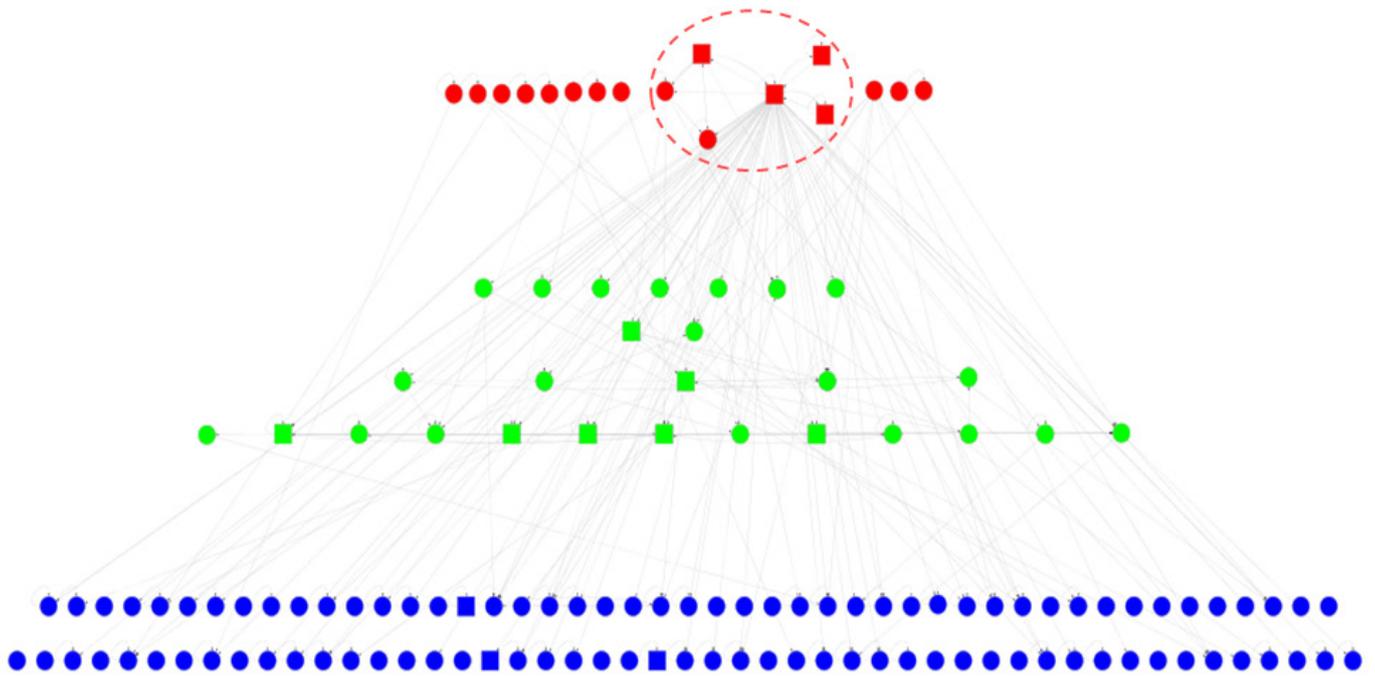

**Figure 1: Hierarchical decomposition of transcriptional regulators into Top, Middle and Bottom levels in the TRN of *B. subtilis* with σ-factors.** The network of transcriptional regulators has a pyramidal structure, where the largest strongly connected component (LSCC) of 6 nodes (encircled with red dashed oval) lies at the Top level of the hierarchy. Transcriptional regulators in the Top, Middle and Bottom levels of hierarchy are shown in Red, Green and Blue, respectively; transcription factors (TFs) are depicted as circles and σ−factors as squares. The network visualization was obtained by using Cytoscape[43].

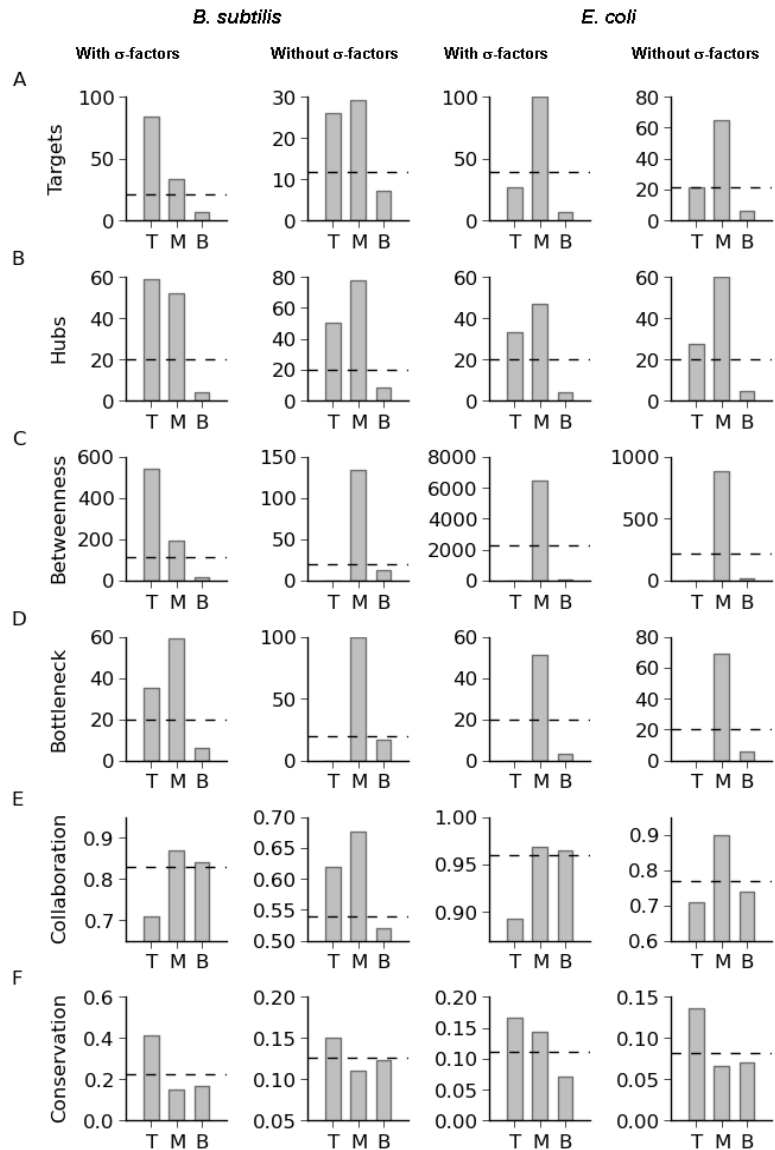

**Figure 2: Enrichment of structural and functional properties in the different levels of hierarchy in the TRNs of *B. subtilis* and *E. coli*.**
(A) Number of targets (out-degree); (B) Distribution of hubs; (C) Betweenness centrality; (D) Distribution of bottlenecks; (E) Degree of collaboration; (F) Evolutionary conservation of transcriptional regulators between *B. subtilis* and *E. coli*. The expected values of the given properties of transcriptional regulators at different levels of hierarchy for randomized networks are shown as dashed black lines.

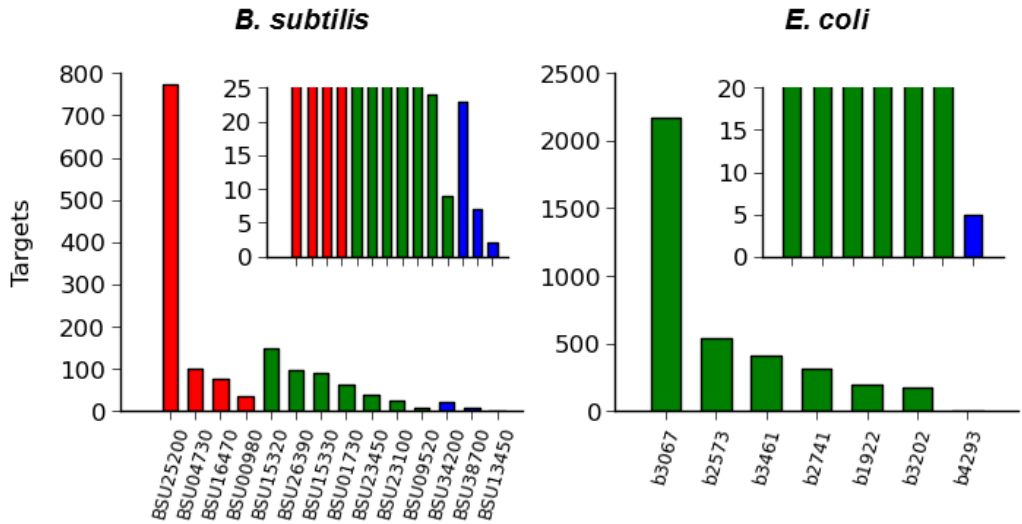

**Figure 3: Out-degree of σ-factors in the TRNs of *B. subtilis* and *E. coli*.** Vertical bars for σ-factors in the Top, Middle and Bottom levels of the hierarchy are shown in Red, Green and Blue, respectively. Insets zoom on to σ-factors with small out-degree. σ-factors in *B. subtilis* are scattered across all three levels of the hierarchy, whereas in *E. coli* almost all the σ-factors are in the Middle level. The σ-factor RpoD (BSU25200 in *B.subtilis* and b3067 in *E. coli*) has the maximum number of targets (out-degree) in both organisms, but occurs in the Top level in *B.subtilis* and in the Middle level in *E. coli*.

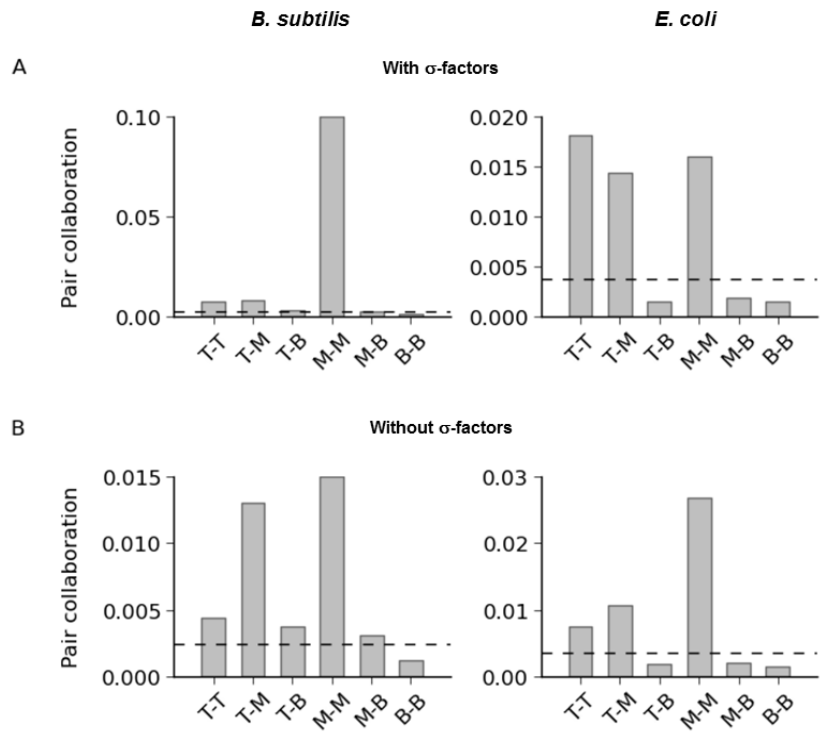

**Figure 4: Extent of intra- and inter-level pair coregulatory partnerships of and between different levels of hierarchy in the TRNs of *B. subtilis* and *E. coli*.** (A) TRNs with σ-factors; (B) TRNs without σ-factors. The average degree of pair collaboration is highest at the Middle-Middle followed by the Top-Middle in all cases, except for the TRN of *E. coli* with σ-factors.

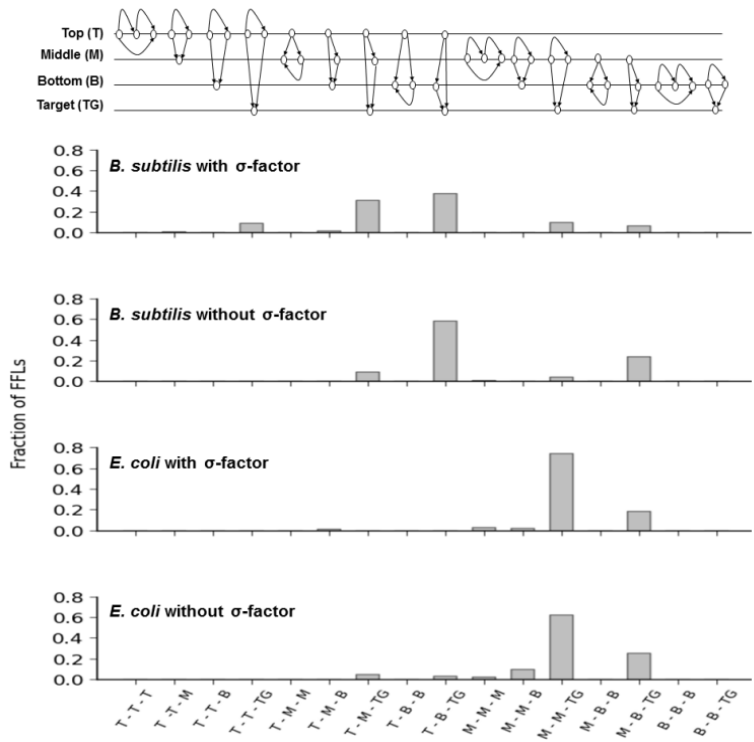

**Figure 5: Composition of feed forward loops (FFLs) for genes from different levels of hierarchy in the TRNs of *B. subtilis* and *E. coli*.** Top level and Bottom level regulators appear more often in FFLs in the TRN of *B. subtilis* while Middle level regulators appear more often in FFLs in the TRN of *E. coli*.

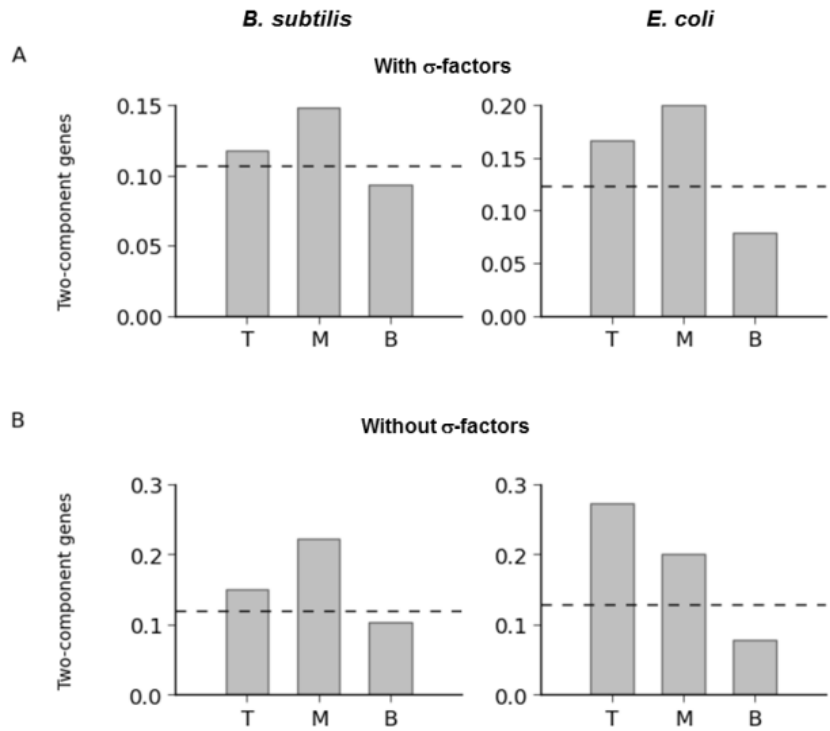

**Figure 6: Distribution of two-component regulatory system genes in different levels of hierarchy in the TRNs of *B. subtilis* and *E. coli*.** (A) TRNs with σ-factors and (B) TRNs without σ-factors. The Middle level is enriched in two-component system transcriptional regulators in both the TRNs of *B. subtilis* and *E. coli*.

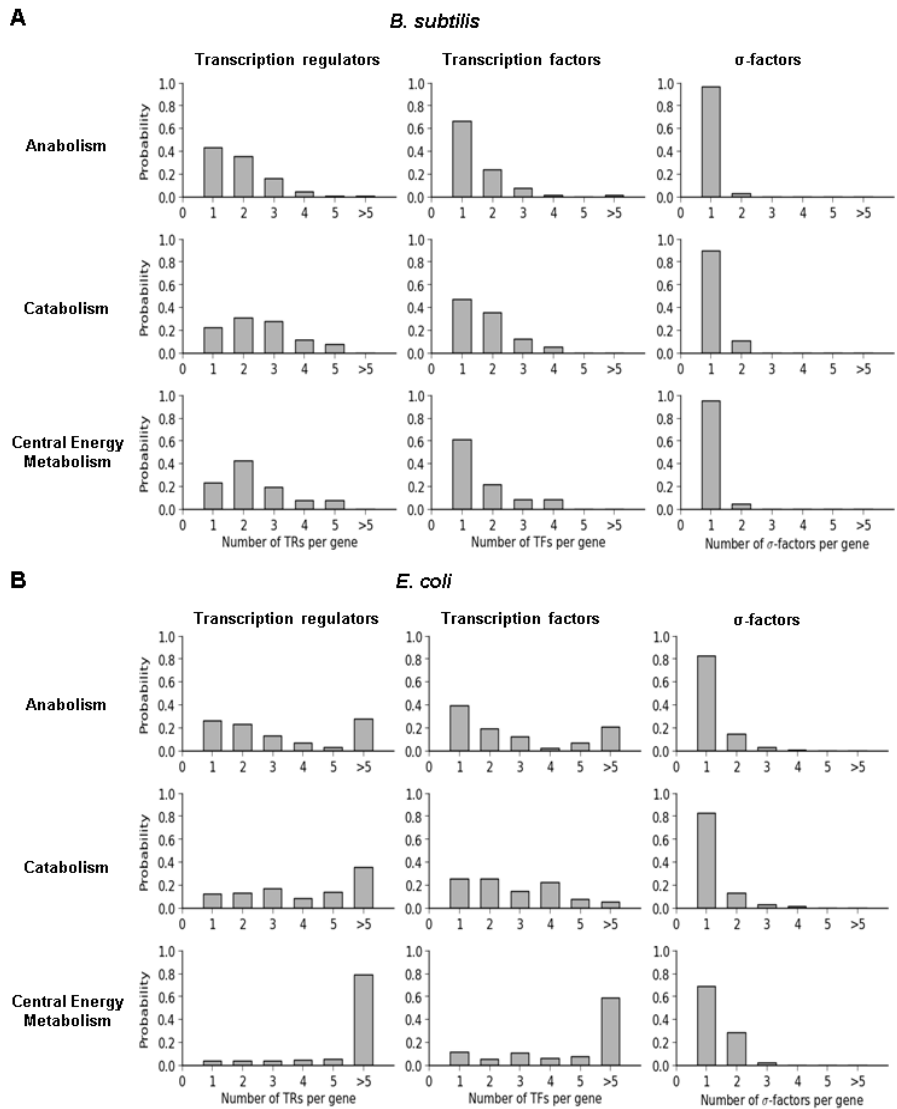

**Figure 7: Regulation of enzymes in distinct metabolic subsystems by transcriptional regulators.** (A) *B. subtilis* and (B) *E. coli*. The regulation of enzymes in distinct metabolic subsystems is shown separately for transcriptional regulators (TRs), transcription factors (TFs) and σ-factors. Anabolic genes are the least regulated ones in both organisms.